\documentclass[12pt]{article}
\usepackage{graphicx}
\usepackage{amsmath}

\input{tcilatex}

\begin{document}

\title{The Calogero equation and Liouville type equations}
\author{Maxim V.Pavlov \\
City University of Hong Kong,\\
83 Tat Chee Avenue, Kowloon, Hong Kong.,\\
e-mail: maxim.pavlov@mtu-net.ru}
\maketitle

Abstract. In this paper we present a two-component generalization of the
C-integrable Calogero equation (see [1]). This system is C-integrable as
well, and moreover we show that the Calogero equation and its two-component
generalization are solvable by a reciprocal transformation to ODE's.
Simultaneously we obtain a generalized Liouville equation (34), determined
by two arbitrary functions of one variable.

\section{Introduction}

F.Calogero in his article ''A solvable nonlinear wave equation'' presented
the nonevolutionary equation

\begin{equation}
u_{xt}=uu_{xx}+\Phi (u_{x}),  \tag{1}
\end{equation}%
with an arbitrary function $\Phi (p)$. Moreover he have found a general
solution, but in some complicated form. Here we suggest a new approach for
description of this solution in more clear form by using suitable reciprocal
transformation. In result we obtain a general solution for Liouville type
equations (see below).

At first we notice that the Calogero equation (1) has a special conservation
law
\begin{equation}
\partial _{t}F(u_{x})=\partial _{x}[uF(u_{x})],  \tag{2}
\end{equation}%
where
\begin{equation}
F(p)=\exp [\int \frac{pdp}{\Phi (p)}].  \tag{3}
\end{equation}%
The corresponding reciprocal transformation
\begin{equation}
dz=F(p)dx+uF(p)dt\text{, \ \ \ \ }dy=dt  \tag{4}
\end{equation}%
yields recalculation of independent variables
\begin{equation}
\partial _{x}=\upsilon \partial _{z}\text{ \ \ \ and \ \ \ }\partial
_{t}=\partial _{y}+u\upsilon \partial _{z},  \tag{5}
\end{equation}%
where $\upsilon =F(p)$. Then the conservation law (2) has form
\begin{equation}
u_{z}=\partial _{y}(-1/\upsilon )  \tag{6}
\end{equation}%
and the Calogero equation (1) transforms into
\begin{equation}
p_{y}=\Phi (p),  \tag{7}
\end{equation}%
where (see (5) and (6))
\begin{equation}
p=u_{x}=\upsilon u_{z}=\partial _{y}\ln \upsilon .  \tag{8}
\end{equation}%
Thus, a solution of the Calogero equation (1) can be presented in implicit
form (see (7)) and found consequently by integration from (7), (8) and from
the inverse reciprocal transformation
\begin{equation}
dx=\frac{1}{F(p)}dz-udy\text{, \ \ \ }dt=dy.  \tag{9}
\end{equation}%
In second section we present couple well-known examples of such integrable
equations. In third section we describe another reciprocal transformation
from the generalized Hunter-Saxton equation into the Liouville equation. In
fourth section we discuss general reciprocal transformations from a
nonlinear PDE more complicated than the Calogero equation (1) into an ODE of
the second order. In last fifth section we construct a two-component
generalization of the Calogero equation (1) and present its solution by
corresponding reciprocal transformation to ODE of the second order. Thus,
the major aim of this article is possibility to demonstrate that some
C-integrable equations (or systems) can be integrated by suitable reciprocal
transformations into corresponding ODE's.

\section{Remarkable examples}

One of them is the generalized Hunter-Saxton equation
\begin{equation}
u_{xt}=uu_{xx}+\varepsilon u_{x}^{2}  \tag{10}
\end{equation}%
was studied in [2], where a general solution was found by characteristic
method for $\varepsilon =1/2$. Exactly this case was also studied by P.Olver
and P.Rosenau in them article [3] from another point of view. Here we
construct a general solution for arbitrary $\varepsilon $. Since $\Phi
(p)=\varepsilon p^{2}$, then
\begin{eqnarray}
F(p) &=&p^{1/\varepsilon }\text{, \ \ \ }p=\frac{1}{A(z)-\varepsilon y}\text{%
,}  \TCItag{11} \\
u &=&B^{\prime }(y)+\int [A(z)-\varepsilon y]^{(1-\varepsilon )/\varepsilon
}dz\text{,}  \notag \\
x &=&-B(y)+\int [A(z)-\varepsilon y]^{1/\varepsilon }dz.  \notag
\end{eqnarray}%
If $\varepsilon =1/2$, then we obtain a general solution of the
Hunter-Saxton equation
\begin{eqnarray}
u &=&B^{\prime }(y)+\int A(z)dz-\frac{1}{2}yz,  \TCItag{12} \\
x &=&-B(y)+\int A^{2}(z)dz-y\int A(z)dz+\frac{1}{4}y^{2}z.  \notag
\end{eqnarray}%
If $\varepsilon =1$, equation (10) allows the reduction
\begin{equation*}
u_{t}=uu_{x}
\end{equation*}%
Its solution is
\begin{equation*}
x=-B(y)+C(z)-yz\text{, \ \ \ \ }u=B^{\prime }(y)+z
\end{equation*}%
It is easy to check that in this case (just one equation of the first order)
$B(y)=\alpha y+\beta $ and finally we obtain the standard solution
\begin{equation*}
(x+\beta )+ut=C(u-\alpha )
\end{equation*}%
At this conference NEED's 2000 in Gukova (Turkey) one of participants
professor Valery Druma demonstrated for author another nonevolutionary
equation appeared in intersection of projective geometry and gravity (see
[4])%
\begin{equation*}
u_{xxt}=uu_{xxx}.
\end{equation*}%
This equation has obvious integral%
\begin{equation*}
u_{xt}=uu_{xx}-\frac{1}{2}u_{x}^{2}+\gamma (t).
\end{equation*}%
For special case $\gamma =0$ this is particular case of the generalized
Hunter-Saxton equation (10).

\section{The Calogero equation and The Liouville equation}

The generalized Hunter-Saxton system (10) has another conservation law in
comparison with (2)
\begin{equation}
\partial _{t}[(u_{xx})^{1/(2\varepsilon +1)}]=\partial
_{x}[u(u_{xx})^{1/(2\varepsilon +1)}].  \tag{13}
\end{equation}%
The corresponding reciprocal transformation
\begin{equation}
dw=qdx+uqdt\text{, \ \ \ \ }d\tau =dt  \tag{14}
\end{equation}%
yields a recalculation of independent variables
\begin{equation}
\partial _{x}=q\partial _{w}\text{, \ \ \ \ }\partial _{t}=\partial _{\tau
}+uq\partial _{w},  \tag{15}
\end{equation}%
where
\begin{equation}
q=(u_{xx})^{1/(2\varepsilon +1)}.  \tag{16}
\end{equation}%
Then the conservation law (13) transforms into
\begin{equation}
u_{w}=-\partial _{\tau }(1/q)  \tag{17}
\end{equation}%
and simultaneously the generalized Hunter-Saxton equation transforms into
\begin{equation}
s_{\tau }=\varepsilon s^{2},  \tag{18}
\end{equation}%
where
\begin{equation}
s=u_{x}=qu_{w}=\partial _{\tau }\ln q.  \tag{19}
\end{equation}%
Thus a general solution of the generalized Hunter-Saxton equation can be
presented in implicit form (see (18)) and found consequently by integration
from (18), (19) and from the inverse reciprocal transformation
\begin{equation}
dx=\frac{1}{q}dw-ud\tau \text{, \ \ \ \ }dt=d\tau .  \tag{20}
\end{equation}%
This solution is
\begin{equation}
x=\int \frac{dw}{B(w)}[A(w)+\varepsilon \tau ]^{1/\varepsilon }+C(\tau )%
\text{, \ \ \ \ }u=-\int \frac{dw}{B(w)}[A(w)+\varepsilon \tau
]^{(1-\varepsilon )/\varepsilon }-C^{\prime }(\tau )\text{,}  \tag{21}
\end{equation}%
where
\begin{equation}
q=B(w)[A(w)+\varepsilon \tau ]^{-1/\varepsilon }\text{, \ \ \ \ \ \ }%
s=-[A(w)+\varepsilon \tau ]^{-1}.  \tag{22}
\end{equation}%
Simultaneously, anyone can recalculate (16)
\begin{equation}
q^{2\varepsilon +1}=u_{xx}=q\partial _{w}(qu_{w}),  \tag{16a}
\end{equation}%
and substitute (19) into (16a) at next step. Then the Hunter-Saxton equation
transforms by reciprocal transformation (14) into the Liouville equation
\begin{equation}
\partial _{w\tau }\ln q=q^{2\varepsilon }.  \tag{23}
\end{equation}%
Substituting (22) into (23) anyone can find well-known general solution of
the Liouville equation
\begin{equation}
q=[A^{\prime }(w)]^{1/2\varepsilon }[A(w)+\varepsilon \tau ]^{-1/\varepsilon
},  \tag{24}
\end{equation}%
where
\begin{equation*}
B(w)=[A^{\prime }(w)]^{1/2\varepsilon }.
\end{equation*}%
Thus finally the general solution of the Hunter-Saxton equation is
\begin{eqnarray}
x &=&\int [A^{\prime }(w)]^{-1/2\varepsilon }[A(w)+\varepsilon \tau
]^{1/\varepsilon }dw+C(\tau )\text{, \ }  \TCItag{21a} \\
u &=&-\int [A^{\prime }(w)]^{-1/2\varepsilon }[A(w)+\varepsilon \tau
]^{(1-\varepsilon )/\varepsilon }dw-C^{\prime }(\tau )\text{.}  \notag
\end{eqnarray}%
If a general solution of the generalized Hunter-Saxton equation depends on
two functions of one variable (see (21a) or (11)), a general solution of the
Liouville equation must depend on two functions of one variable too. It is
easy to reconstruct by using the obvious symmetry of the Liouville equation $%
\tau \rightarrow R(\tau )$, \ $q\rightarrow q[R^{\prime }(\tau
)]^{1/2\varepsilon }$ (see (24))
\begin{equation}
q=[A^{\prime }(w)R^{\prime }(\tau )]^{1/2\varepsilon }[A(w)+\varepsilon
R(\tau )]^{-1/\varepsilon }.  \tag{24a}
\end{equation}%
The particular case ($\varepsilon =1/2$) was studied in [5] for
high-frequency limit of Camassa-Holm equation. However, here we showed that
the generalized Hunter-Saxton equation transformable into the Liouville
equation. Thus, in this section we described relationship between the
Hunter-Saxton equation and the Liouville equation, simultaneously we
integrated as the Hunter-Saxton equation as the Liouville equation. However,
in next section we present general scheme for integrability of the Calogero
equation and we will show, that by this way the Calogero equation can be
transform into more general equation than the Liouville equation, and
simultaneously the Calogero equation can be integrated again by reciprocal
transformation into ODE.

\section{A General Case}

We suggest that the nonevolutionary equation
\begin{equation}
u_{xt}=\varphi (u,u_{x},u_{xx})  \tag{25}
\end{equation}%
has the special conservation law
\begin{equation}
\partial _{t}f(u_{x},u_{xx})=\partial _{x}[uf(u_{x},u_{xx})].  \tag{26}
\end{equation}%
Then right-hand side of (25) is more determined
\begin{equation}
u_{xt}=uu_{xx}+\psi (u,u_{x}),  \tag{25a}
\end{equation}%
where from compatibility condition of (25) and (26) we have
\begin{equation}
\psi \frac{\partial f}{\partial u_{x}}+[\frac{\partial \psi }{\partial u}%
u_{x}+(u_{x}+\frac{\partial \psi }{\partial u_{x}})u_{xx}]\frac{\partial f}{%
\partial u_{xx}}=fu_{x}.  \tag{27}
\end{equation}%
If we introduce the reciprocal transformation
\begin{equation}
dz=wdx+uwdt\text{, \ \ \ }dy=dt  \tag{28}
\end{equation}%
for (25a), where $w=f(u_{x},u_{xx})$, then we obtain from (25a)
\begin{equation}
(wu_{z})_{y}=\psi (u,wu_{z})  \tag{29}
\end{equation}%
and
\begin{equation}
u_{z}=w^{-2}w_{y}.  \tag{30}
\end{equation}%
Thus, we finally obtain \textit{ordinary differential equation}
\begin{equation}
\partial _{y}^{2}\ln w=\psi (u,\partial _{y}\ln w)  \tag{29a}
\end{equation}%
and
\begin{equation}
w=f(\partial _{y}\ln w,w\partial _{yz}\ln w).  \tag{31}
\end{equation}%
However, the ODE (29a) can be integrated just if $\psi =$ $\psi (u_{x})$.
Then the equation (27)
\begin{equation}
\psi \frac{\partial f}{\partial u_{x}}+[u_{x}+\psi ^{\prime }(u_{x})]u_{xx}%
\frac{\partial f}{\partial u_{xx}}=fu_{x}.  \tag{27a}
\end{equation}%
can be integrated immediately%
\begin{equation}
f=\beta (u_{x})\zeta (\frac{\beta (u_{x})\psi (u_{x})}{u_{xx}}),  \tag{32}
\end{equation}%
where
\begin{equation}
\beta (\tau )=\exp [\int \frac{\tau d\tau }{\psi (\tau )}]  \tag{33}
\end{equation}%
and $\zeta (\upsilon )$ is an arbitrary function. Thus the Calogero equation
(1) by generalized conservation law (26) (see reciprocal transformation
(28)) transforms into an \textbf{ordinary differential equation} (see (29)
and (29a))
\begin{equation}
s_{yy}=\psi (s_{y})  \tag{29b}
\end{equation}%
and simultaneously into a hyperbolic equation (see (31))%
\begin{equation}
s_{yz}=\psi (s_{y})\alpha (\beta (s_{y})e^{-s}),  \tag{34}
\end{equation}%
where
\begin{equation}
s=\ln w\text{, \ \ \ }\zeta (\frac{1}{\tau \alpha (\tau )})=\tau \text{.}
\tag{35}
\end{equation}%
Thus, we obtain following beautiful fact: every hyperbolic equation (34)
with two arbitrary functions $\psi (\tau )$ and $\alpha (\tau )$ is
C-integrable (see (29b)), if $\beta (\tau )$ is satisfying to (33). From
this point of view this is natural generalization of the Liouville equation
(23).

\section{Two-component Generalization}

The Calogero equation (1) allows the two-component generalization
\begin{equation}
\eta _{t}=\partial _{x}(u\eta )\text{, \ \ \ }u_{xt}=uu_{xx}+\psi (\eta
,u_{x}).  \tag{36}
\end{equation}%
By application of the reciprocal transformation
\begin{equation}
dz=\eta dx+u\eta dt\text{, \ \ \ }dy=dt  \tag{37}
\end{equation}%
this system (36) transforms into \textbf{ordinary differential equation }%
(see (29b) for comparison)
\begin{equation}
s_{yy}=\psi (e^{s},s_{y}),  \tag{38}
\end{equation}%
where $s=\ln \eta $. Integrability of this system (36) looks simpler than of
the Calogero equation (1), but the ODE (38) is more complicated and
reduction from (36) into (1) is not so obvious $\eta =\eta (u_{x})$, where
function $\eta $ is solution of another ODE%
\begin{equation*}
\psi (\eta ,g)d\eta =\eta gdg.
\end{equation*}%
In next article [6] we show that one particular case of this two-component
generalization (36), when $\psi =\frac{1}{2}(\eta ^{2}+u_{x}^{2})$ (motion
of dark matter in Universe, see [7]) is closely related with nonlinear
Shrodinger -- Maxwell-Bloch hierarchy. In this case, system (36) has
infinite set of local Hamiltonian structures, commuting flows and
conservation laws and can be written in Monge-Ampere form. Alternative
interpretation of this special case was presented in [8] from nonlinear
optics point of view.

This article partially was supported by the City University of Hong Kong
(grant \#7001041) and the Research Grants Council of Hong Kong (grant
\#9040466). Also the author is grateful to RFBR (grants \#00-01-00210 and
\#00-01-00366) and thanks City University of Hong Kong for hospitality.

\section{References}

[1]. Calogero, F. A solvable nonlinear wave equation. Stud.Appl.Math.70
(1984), no 3,

189-199.

[2]. Hunter, John K.; Saxton, Ralph, Dynamics of director fields. SIAM J.
Appl.

Math. 51 (1991), no. 6, 1498--1521.

[3]. P.Olver, P.Rosenau, Tri-Hamiltonian duality between solitons and
solitary wave

solutions having compact support. Phys.Rev.E (3), 53 (1996), No.2, 1900-1906.

[4]. Dryuma, Valerii, On initial values problem in theory of the second
order ordinary

differential equations. Proceedings of the Workshop on Nonlinearity,
Integrability and

All That: Twenty Years after NEEDS '79 (Gallipoli, 1999), 109--116, World
Sci.

Publishing, River Edge, NJ, 2000.

[5]. Dai, Hui-Hui; Pavlov, Maxim Transformations for the Camassa-Holm
equation,

its high-frequency limit and the Sinh-Gordon equation. J. Phys. Soc. Japan 67

(1998), no. 11, 3655--3657.

[6]. Pavlov, Gurevich-Zybin system, will be published.

[7]. A.V.Gurevich, K.P.Zybin, ''Nondissipative gravitational turbulence,
Soviet

Phys.JETP 67 (1) (1988) 1957; A.V.Gurevich, K.P.Zybin, ''Large-scale\
structure of

the Universe.\ Analytic theory'', Soviet Phys.Usp. 38 (7) (1995), 687.

[8]. A.M.Kamchatnov, M.V. Pavlov, Title: A new C-integrable limit of SHG

equations, nlin.SI/0012014, will be published in Journal of Physics A, Math.
and Gen.

\end{document}